\documentclass[preprint,12pt]{elsarticle}

\usepackage[utf8]{inputenc}
\usepackage{amssymb}
\usepackage{times}
\usepackage{units}
\usepackage{fullpage}
\usepackage{array}
\usepackage{tabularx, booktabs}
\usepackage[colorlinks=true, linkcolor=blue, citecolor=blue]{hyperref}
\usepackage{fancyhdr}
\usepackage{multirow}
\usepackage{upgreek}
\usepackage{epstopdf}
\usepackage[usenames,dvipsnames]{xcolor}
\usepackage{graphicx}
\usepackage{caption}
\usepackage{wasysym}
\usepackage[labelformat=simple]{subcaption}

\usepackage{float}
\usepackage{lineno}

\biboptions{sort&compress}
\journal{Composite Structures}

\newcommand{\rev}[1]{{\color{black}#1}}

\makeatletter
\def\ps@pprintTitle{
	\let\@oddhead\@empty
	\let\@evenhead\@empty
	\def\@oddfoot{{\small© 2018. This manuscript version is made available under the \href{http://creativecommons.org/licenses/by-nc-nd/4.0/}{ CC-BY-NC-ND 4.0 license}.}\hfill{}}%
	\let\@evenfoot\@oddfoot
}
\makeatother

\begin{document}

\begin{frontmatter}



\author[ctu]{V.~Ne\v{z}erka}
\author[ctu]{M.~Somr}
\author[ctu]{T.~Janda}
\author[ctu]{J.~Vorel}
\author[ctu]{M.~Do\v{s}k\'{a}\v{r}}
\author[ctu]{J.~Anto\v{s}}
\author[ctu]{J.~Zeman}
\author[ctu]{J.~Nov\'{a}k\corref{cor1}}
\cortext[cor1]{Corresponding author}
\ead{novakja@fsv.cvut.cz}

\address[ctu]{Faculty of Civil Engineering, Czech Technical University in Prague, Th\'{a}kurova~7, 166~29 Praha~6, Czech Republic}

\title{A jigsaw puzzle metamaterial concept\tnoteref{t1}}
\tnotetext[t1]{Author's post-print version of the article manuscript published in \mbox{\textit{Composite Structures}}\\\href{https://doi.org/10.1016/j.compstruct.2018.06.015}{DOI: 10.1016/j.compstruct.2018.06.015}.}

\begin{abstract}

A concept of a planar modular mechanical metamaterial inspired by the principle of local adaptivity is proposed. The metamaterial consists of identical pieces similar to jigsaw puzzle tiles. Their rotation within assembly provides a substantial flexibility in terms of structural behavior, whereas mechanical interlocks enable reassembly. The tile design with a diagonal elliptical opening allows us to vary elastic properties from stiff to compliant, with positive, zero, or negative Poisson's ratio. The outcomes of experimental testing on additively manufactured specimens confirm that the assembly properties can be accurately designed using optimization approaches with finite element analysis at heart.

\end{abstract}

\end{frontmatter}

%\linenumbers


\section{Introduction} \label{sec:intro}

%\noteMD{Myslim, ze bychom meli zminit Cheung, K.C., Gershenfeld, N., 2013. Reversibly Assembled Cellular Composite Materials. Science 341, 1219–1221. https://doi.org/10.1126/science.1240889. Maji totiz pristup, jak vytvorit DEMONTOVATELNOU strukturu z jednoho prvku.}
%\noteMD{Zvazte, zda nedat odkaz na Malik, I.A., Mirkhalaf, M., Barthelat, F., 2017. Bio-inspired “jigsaw”-like interlocking sutures: Modeling, optimization, 3D printing and testing. Journal of the Mechanics and Physics of Solids 102, 224–238. https://doi.org/10.1016/j.jmps.2017.03.003, kde se zabyvaji pull-out a celkove mechanickou odezvou puzzle interlocku.}
Nature efficiently distributes material and designs optimal structure across scales with respect to anticipated loading. Bone tissue is a perfect example of such a multiscale hierarchical structure in which the organization of mineral nanoparticles and collagen microfibrils governs the bone's elastic properties~\cite{Granke_2013}. Here we propose to mimic such local adaptivity by material design similarly to a way children assemble images decomposed into jigsaw puzzle pieces. We create a planar block of mechanical metamaterial~\cite{Wegener_2013} with required overall elastic stiffness emanating from the geometry and composition of mesostructural units, similar to the concepts proposed in, e.g.,~\cite{Wu_2015, Schumacher_2015, Mousanezhad_2017,Cheung_2013}. 
\begin{figure}[!htp]
\centering
%% \begin{tabular}{cc}
%% 	\includegraphics[width=0.3\textwidth]{FIG2a.jpg} & \includegraphics[width=0.313\textwidth]{FIG2b.jpg}\\
%% 	(a) & (b)
%% \end{tabular}
\includegraphics[width=0.5\textwidth]{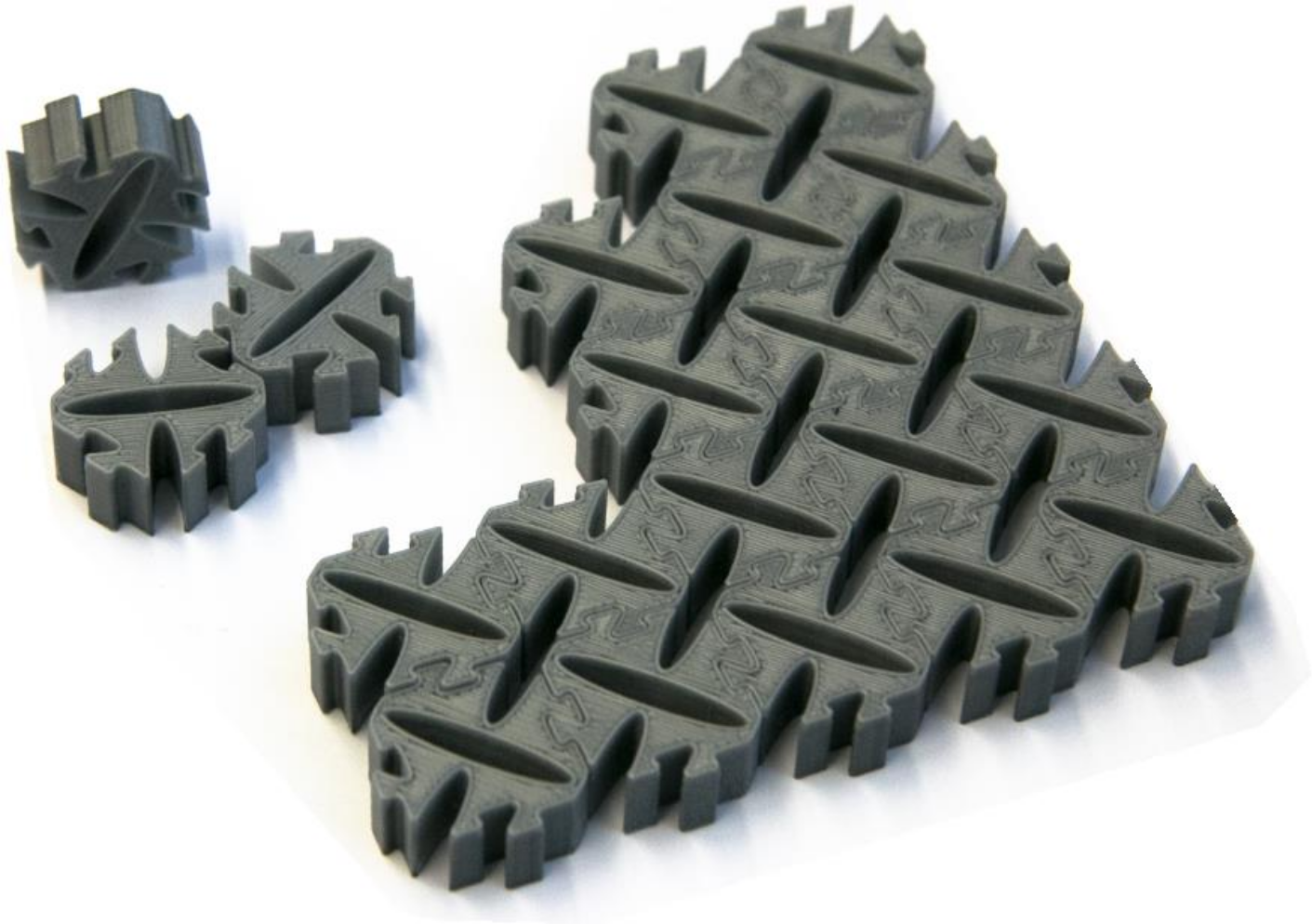}
\caption{\rev{Additively manufactured jigsaw puzzle tiles and partly disassembled auxetic aggregate with effective Poisson's ratio equal to $-0.55$.}}
\label{fig:assemblies}
\end{figure}
In particular, the assemblies are composed of mutually rotated identical tiles with elliptic openings, Figure~\ref{fig:assemblies}, as proposed by Taylor et al.~\cite{Taylor_2014}.
% \rev{or authors themselves e.g. in~\cite{nevzerka2017additively}}.
In addition, the subscale, tile-level, geometry design also allows us to tune Poisson's ratio~\cite{Overvelde_2012, Overvelde_2014} and thus fabricate materials exhibiting conventional or auxetic~\cite{Lakes_1987, Ai_2018} behavior. \rev{The tiles in assemblies can be connected through jigsaw puzzle interlocks~\cite{Malik_2017}, or the modularity concept can serve for optimization and design purposes and the conglomerates can be printed out as monolithic pieces~\cite{Schumacher_2015}. In this work,} the interlocks are not glued together, they utilize only frictional forces to meet the requirement on reasonably stiff contact and non-destructive reassembly. \rev{In order to obtain comprehensive full-field information on strain and displacement fields, a digital image correlation (DIC)~\cite{Peters_1982, Blaber_2015} analysis is employed throughout the study.}

\section{Modular concept} \label{sec:fabrication}

\begin{figure}[!htp]
\centering
\begin{tabular}{cc}
	\includegraphics[width=0.27\textwidth]{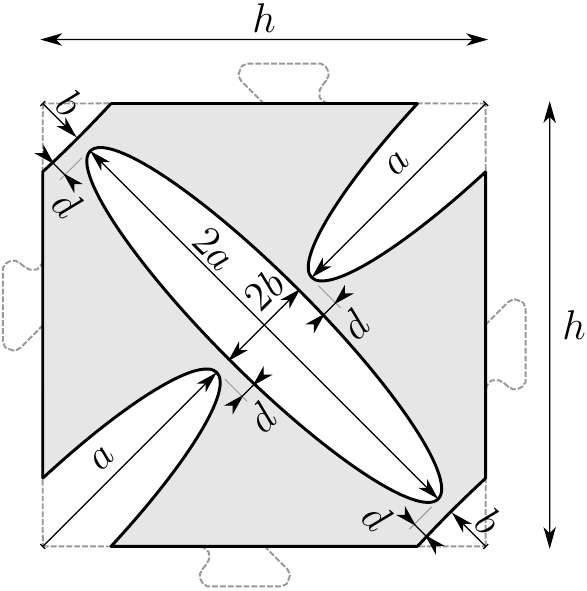} & \includegraphics[width=0.27\textwidth]{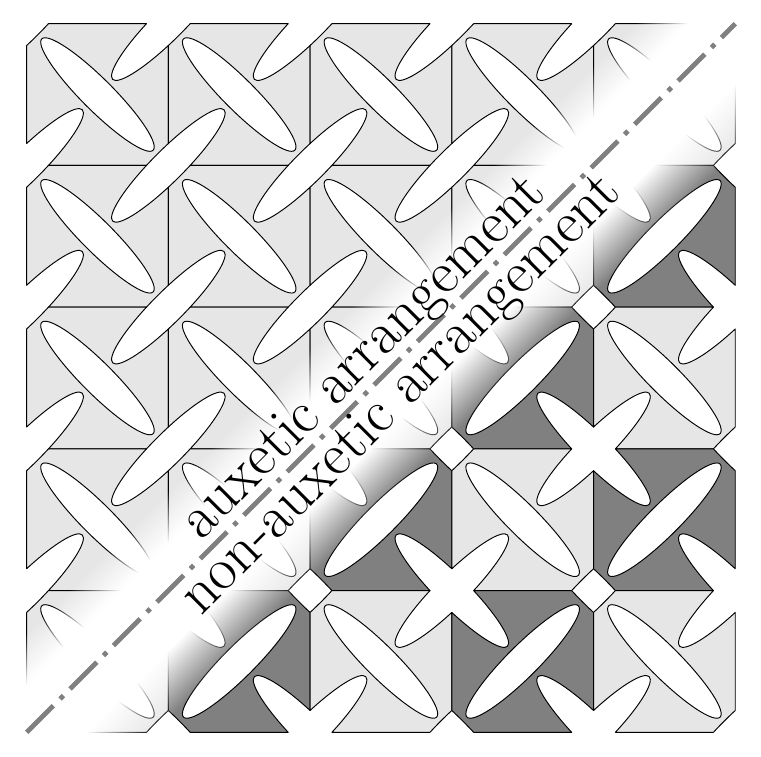}\\
	(a) & (b)
\end{tabular}
\caption{\rev{(a) geometry of single jigsaw puzzle tile with interlocks outlined by dotted contours. (b) regular assemblies into either auxetic or non-auxetic arrangements. Highlighted tiles in the latter arrangement are rotated by $90^\circ$.
Particular dimensions of individual parameters of single tile are derived from the tile edge length $h$, here set to $20mm$, as follows: major semiaxis $a=h/2$, minor semiaxis $b=h/10$, yielding ligament thickness $d \approx h/12$. 
%Tiles were manufactured with $h=20~\textrm{mm}$, $a=10.5~\textrm{mm}$, $b=2~\textrm{mm}$ and $d\approx1.64~\textrm{mm}$.
%Note, magnitude of $d$ be derived by trial-and-error method with the objective of favoring deformations at ligaments rather then mechanical interlocks.
}}
\label{fig:puc_geometry}
\end{figure}

The main ingredient of the proposed concept are square tiles of a jigsaw puzzle-like shape, as shown in \rev{Figure~\ref{fig:puc_geometry}(a)}. Besides the obvious assembly into periodic arrangements\rev{, Figure~\ref{fig:puc_geometry}(b)}, the rotational symmetry of the interlocks allows us to perturb the regular arrangement leading to locally adjusted elastic behavior~\cite{Coulais_2016}.

\section{Metamaterial assembly} \label{sec:metamaterial}

\begin{figure}[htp]
  \centering	
  \begin{tabular}{cccc}
    \includegraphics[height=0.21\textheight]{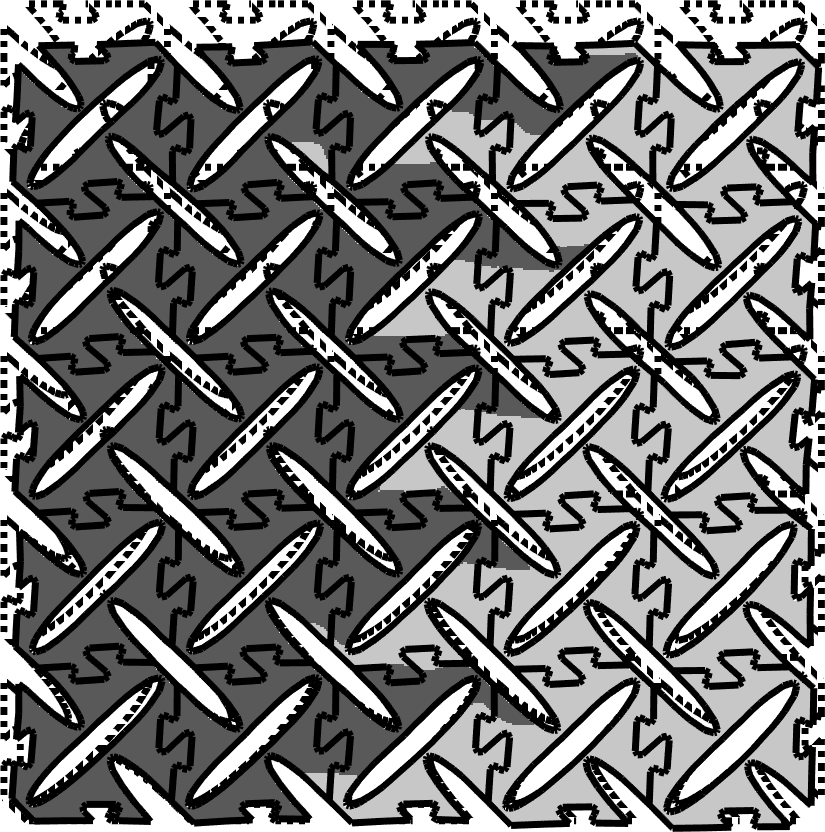} & \includegraphics[height=0.21\textheight]{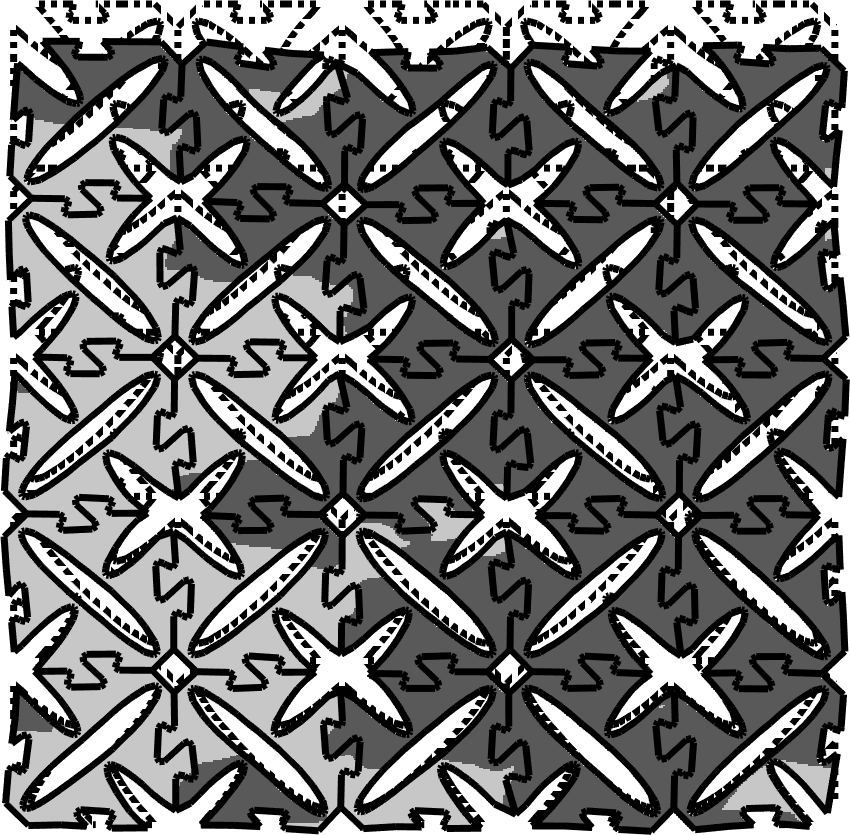} & \includegraphics[height=0.21\textheight]{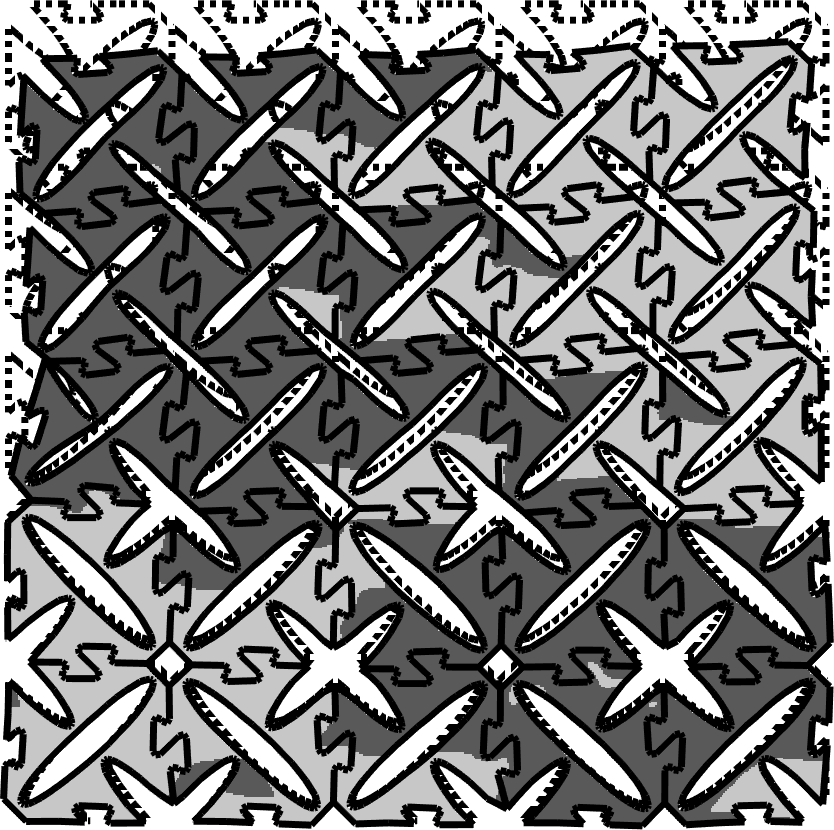} & \includegraphics[height=0.21\textheight]{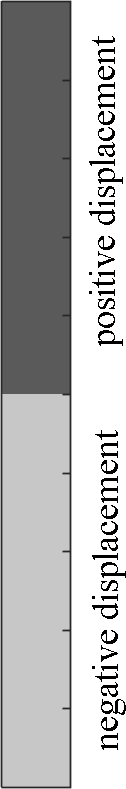}\\
    (a) & (b) &(c) &
\end{tabular}
  \caption{\rev{Specimen deformations with binary representation of horizontal displacements, obtained by DIC, under uniaxial compression applied to (a) auxetic, (b) non-auxetic, and (c) mixed assemblies. Positive displacements are measured according to $x$ and $y$ axes pointing right- and up-wards from the specimen center.}}
  \label{fig:poissonHorizDisplacement_BW}
\end{figure}

To illustrate the adaptivity through local tile rotations, assemblies consisting of $5\times5$ tiles were arranged to exhibit auxetic, non-auxetic, and mixed behavior and subjected to a displacement-controlled compression applied \rev{at the top edge/facet of the specimen}. 
\begin{figure}[!htp]
  \centering	
  \setlength{\tabcolsep}{0pt}
  \begin{tabular}{>{\centering}m{0.30\textwidth} >{\centering}m{0.30\textwidth} >{\centering}m{0.305\textwidth} }
  	\multicolumn{3}{c}{\includegraphics[width=1.00\textwidth]{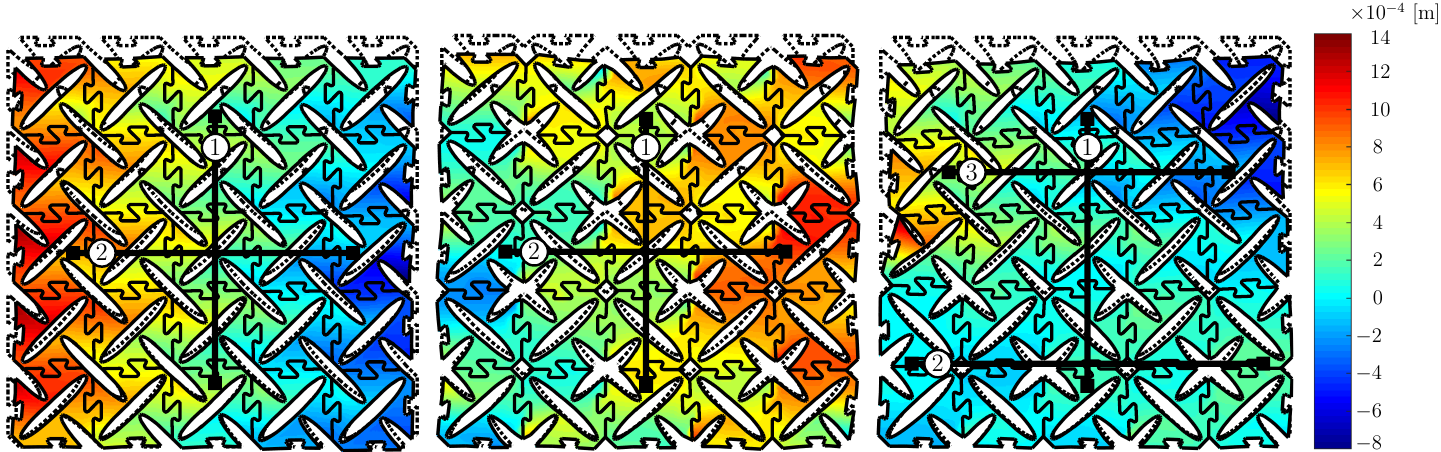}} \\
  	(a) & (b) & (c)
  \end{tabular}  
  \caption{\rev{Specimen deformation (original shape indicated by dotted contours) and horizontal displacement distribution maps, obtained by DIC, under uniaxial compression applied to (a) auxetic, (b) non-auxetic, and (c) mixed assemblies. Virtual extensometers are indicated with boldface lines. Positive displacements are measured according to $x$ and $y$ axes pointing right- and up-wards from the specimen center.}}
  \label{fig:poissonHorizDisplacement}
\end{figure}
First, we explored two regular cases: (i) an auxetic arrangement and (ii) a non-auxetic arrangement, see Figures~\ref{fig:puc_geometry}(b) and~\ref{fig:poissonHorizDisplacement}(a,b). \rev{In addition to the two configurations illustrated also in Figure~\ref{fig:poissonHorizDisplacement_BW}(a,b), a mixed arrangement (iii) supposed to combine the behavior of previous assemblies was also experimentally tested, Figures~\ref{fig:poissonHorizDisplacement}(c) and~\ref{fig:poissonHorizDisplacement_BW}(c).}

\begin{figure}[htp]
  \centering  
  \begin{tabular}{cc}
  	\includegraphics[width=0.475\textwidth]{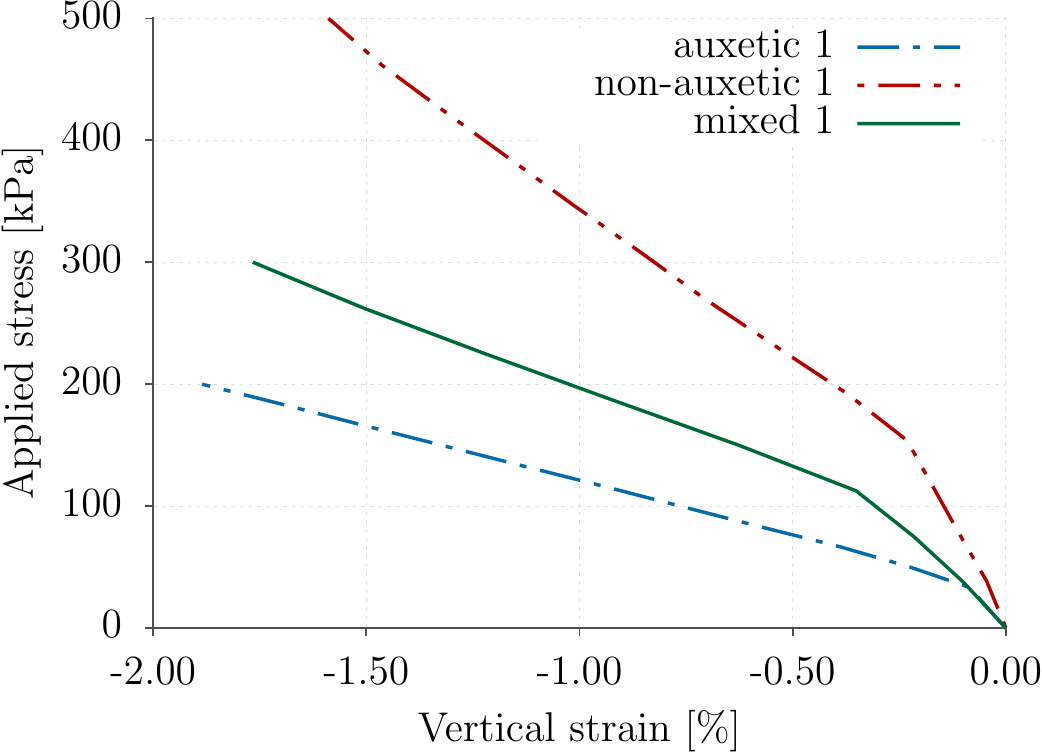} & \includegraphics[width=0.475\textwidth]{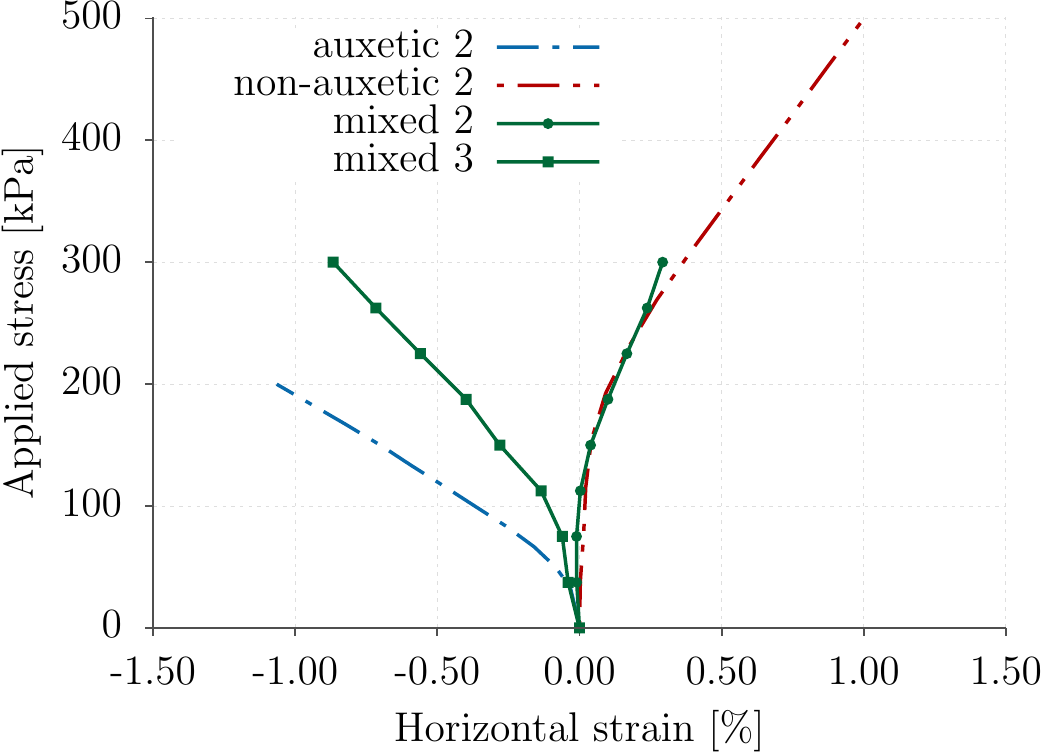}\\
  	(a) & (b)
  \end{tabular}
  \caption{\rev{Effective stress-strain curves for (a) vertical and (b) horizontal extensometers placed in auxetic, non-auxetic, and mixed assemblies according to Figure~\ref{fig:poissonHorizDisplacement}.}}
  \label{fig:F-d_diagram}
\end{figure}

The regular zigzag arrangement of ellipses (i) yielded the smallest value for both elastic constants, Young's modulus and Poisson's ratio. As expected, the specimen contracted in horizontal direction when compressed vertically, Figure~\ref{fig:F-d_diagram}(b), \rev{demonstrating the desired auxetic response}. On the other hand, the configuration (ii) with star-like openings exhibited twice as stiff response, see Figure~\ref{fig:F-d_diagram}(a), and Poisson's ratio exceeding the one of the virgin polylactic acid (PLA) filament. The deformation of the mixed arrangement (iii) clearly proves that Poisson's ratio can be tuned locally, the bottom layer expands horizontally, while the upper part exhibits the auxetic behavior\rev{, Figure~\ref{fig:F-d_diagram}(b)}. This is also \rev{evident from} the \rev{binary patterns} in Figure~\ref{fig:poissonHorizDisplacement_BW}(c).

\section{Customized assemblies}

\begin{figure}[htp]
\centering
  \begin{tabular}{>{\centering}m{0.30\textwidth} >{\centering}m{0.28\textwidth} >{\centering}m{0.21\textwidth} }
  	\multicolumn{3}{c}{\def\svgwidth{0.99\linewidth} \begin{scriptsize} 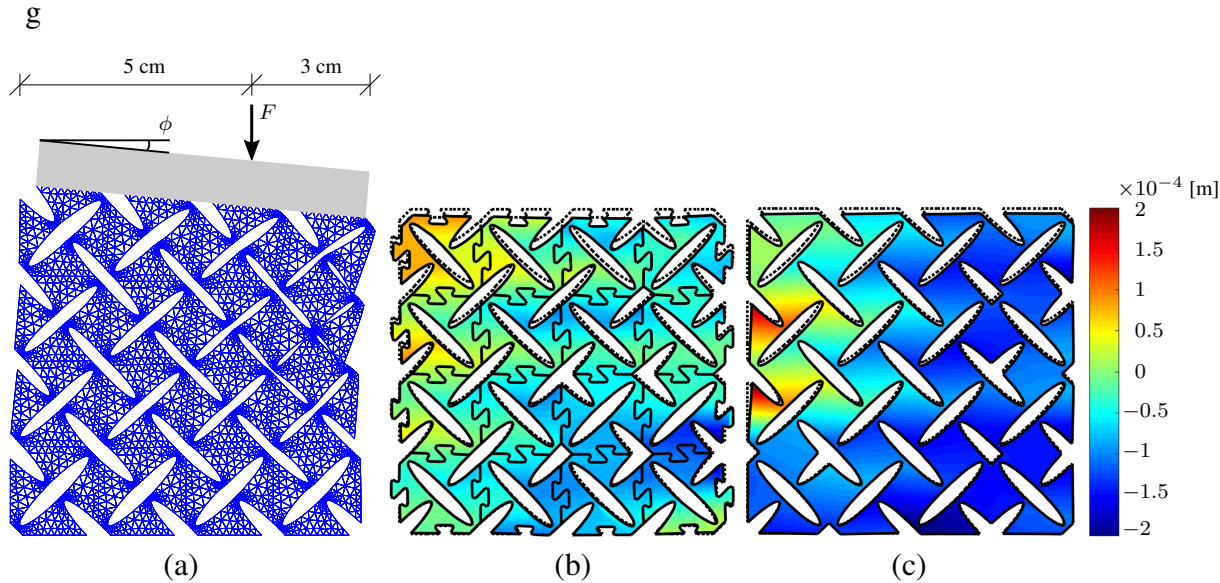 \end{scriptsize}} \\
  	(a) & (b) & (c)
  \end{tabular} 
  \caption{\rev{(a) eccentric loading scheme and the measured tilt angle $\phi$. Experimentally obtained horizontal displacements (original shape indicated by dotted contours) obtained by DIC measurements on specimens designed to compensate eccentric loading: (b) assembly with imperfect mechanical interlocks and (c) without interfaces.}}
  \label{fig:mesh-opt_assemblies}
\end{figure}

The merit of the proposed material system goes beyond many periodic metamaterial designs~\cite{Florijn_2014, Coulais_2017, Zheng_2014} because of its modularity~\cite{Wu_2015, Yang_2017, Mousanezhad_2017} and inherent aperiodicity~\cite{Coulais_2016}. Rotation of a few tiles by $90^{\circ}$ within an assembly \rev{significantly} changes the response to loading. The assembly plan can be adjusted with respect to anticipated loading and specific requirements on both local and global behavior~\cite{Tammas_Williams_2017, Ma_2018}. To demonstrate this \rev{feature}, two modular assemblies (fabricated without interfaces and with imperfect mechanical interlocks) of $4\times4$ tiles were optimized employing a finite element analysis to eliminate tilt angle $\phi$ due to eccentric loading, as \rev{indicated} in Figure~\ref{fig:mesh-opt_assemblies}(a).

%\rev{Two variants of the customized assemblies were considered. First, interfaces between individual tiles were considered perfectly rigid as for both, finite element based optimal design and experimental validation. Unlike the aggregates with weaker mechanical interlocks allowing for a non-destructive reassembly, and such that were considered next, the rigid connections are way easier to handle within the numerical framework, and are foreseen to be better suited for heavily loaded structures. Tested assemblies were manufactured as monoliths, but with the modular design strategy with compatible meshes without contacts in mind. On the other hand, imperfect contacts within aggregates with mechanical interlocks were introduced into the finite element model by means of contact elements and the entire procedure from design to fabrication sustained fully modular.} 

\rev{Two variants of the customized assemblies were considered. First, interfaces between individual tiles were considered perfectly rigid as these are way easier to handle within the numerical framework.
%, and are foreseen to be better suited for heavily loaded structures. 
Tested assemblies were then manufactured as monoliths, but with the modular design strategy with compatible meshes without contacts in mind. 
Second, imperfect contacts within aggregates with mechanical interlocks were introduced into the finite element model by means of contact elements and the entire procedure from design to fabrication sustained fully modular.
}

The tilt angle $\phi$ of the upper edge across all admissible assembly combinations ranged from \rev{$-0.100^{\circ}$ to $1.912^{\circ}$ for solid assemblies and from $-0.108^{\circ}$ to $2.249^{\circ}$ for those with imperfect contacts}. The optimal assemblies that were supposed to yield zero tilt angle, both solid and with mechanical interlocks, were additively manufactured and tested to validate the numerical model. As expected, the optimum assembly plan with mechanical interlocks was completely different from the one with fixed contacts, as shown in Figure~\ref{fig:mesh-opt_assemblies}(b,c), demonstrating the key role of connections. The experimental tilt angle was measured from the displacement distribution maps using virtual extensometers within the DIC framework. The optimum solid and mechanically connected assemblies yielded \rev{$\phi = -0.00022^{\circ}$ and $\phi = 0.01^{\circ}$}, respectively.

%% \begin{figure}[htp]
%% \centering
%%   \def\svgwidth{0.99\linewidth}
%%   \begin{scriptsize}
%%   \input{FIG4.pdf_tex}
%%   \end{scriptsize}
g%%   \caption{Loading scheme of an assembly and the measured tilt angle (a); experimentally obtained deformations (original shape indicated by dotted contours) and maps of horizontal displacements, obtained by DIC, on specimens designed to compensate eccentric loading: assemblies with imperfect mechanical interlocks (b) and without interfaces (c).}
%%   \label{fig:mesh-opt_assemblies}
%% \end{figure}

\section{Materials and methods}

\subsection{Optimization of customized assemblies.}
In the spirit of the explicit complete enumeration method, the computation of all $2^{16} = 65,403$ combinations of $4\times4$ assemblies with the selection of the best solution was done using in-house scripts developed in MATLAB\footnote{The MathWorks, Inc., Matlab release 2015a, \url{http://www.mathworks.com/products/matlab/}, Natick, Massachusetts, U.S.} software linked to ANSYS\footnote{{ANSYS} Mechanical, Academic research, release 16.2, \url{http://www.ansys.com/}, Pennsylvania, U.S.}. These were discretized by $6,184$ linear triangular (assemblies without interfaces, see Figure~\ref{fig:mesh-opt_assemblies}(a)) and $13,322$ isoparametric four-node quadrilateral (assemblies with imperfect interlocks) finite elements PLANE182 under plane strain assumptions. Properties of the contact elements, used for simulations of imperfect interfaces between tiles, were calibrated based on experimental measurements and inverse calculations on pairs of interlocked tiles. In particular, the augmented Lagrangian contact algorithm was employed with the automatic update, using the default values for interfacial contact stiffness parameters provided by the ANSYS program.

\subsection{Fabrication and materials.}
The $2\times2\times1$~cm tiles for experimental verification were additively manufactured from PLA filament with Young's modulus of about $1,800$~MPa and Poisson's ratio equal to $0.3$~\cite{Tymrak_2014}, using Prusa~i3 3D~printer.

\subsection{Mechanical testing and data acquisition.}
The aim of the experimental analysis was to measure macroscopic response of the tested assemblies to displacement-controlled compression. The experiments were carried out using LabTest 4.100SP1 testing machine. The $5\times5$ assemblies were loaded at rate $1.2$--$2.0$~mm/min until the displacement reached $1.6$~mm, while the optimized $4\times4$ assemblies were loaded at the same rate until the eccentrically applied force (see Figure~\ref{fig:mesh-opt_assemblies}) reached $200$~N. In both cases, a steel plate of $1$~cm in thickness was used to distribute the loading over the upper edge of the assemblies.

The analyzed images were taken in $2$-second intervals by a high-definition camera Canon~EOS~70D in uncompressed format (.raw), yielding a resolution of $12$~px/mm. The effect of lens distortion was minimized by setting the focal length to $55$~mm. A non-commercial open-source software Ncorr~\cite{Blaber_2015} was used to evaluate the fields, and a postprocessing of the DIC data was accomplished using an in-house software Ncorr\_post\footnote{V.~Ne\v{z}erka, Ncorr\_post v2.0: DIC Post-Processing Tool, \url{http://mech.fsv.cvut.cz/~nezerka/DIC/index.htm}, Czech Technical University in Prague.}. Virtual extensometers were placed at distinct locations, as depicted in Figure~\ref{fig:poissonHorizDisplacement}, in order to quantify the relative vertical and horizontal displacements of the entire assemblies.

\section{Conclusions}

\begin{figure}[!htp]
  \centering	
  \begin{tabular}{ccccc}
    \includegraphics[width=0.17\textwidth]{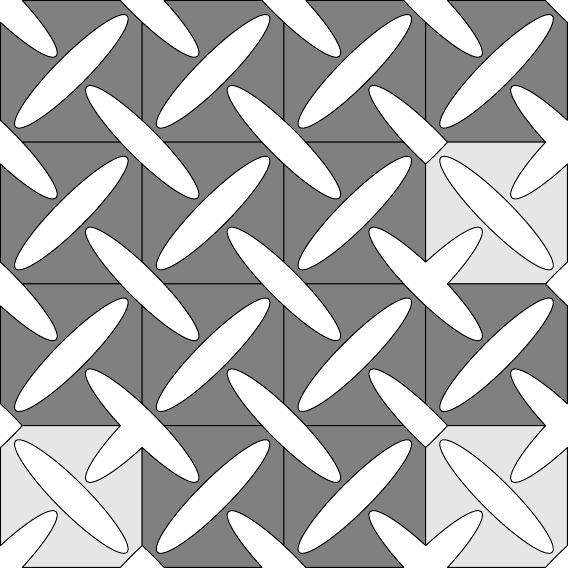} &\includegraphics[width=0.17\textwidth]{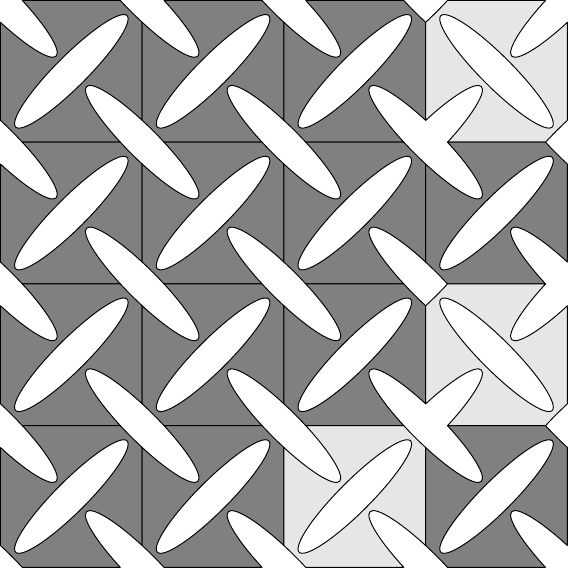}\\
    (a) & (b)
\end{tabular}
  \caption{\rev{Solid assemblies for tilt angles, (a) $\phi=0.00022^\circ$, (b) $\phi=0.00029^\circ$. Highlighted tiles are rotated by $90^\circ$ with respect to the setting adopted in Figure~\ref{fig:puc_geometry}(a).}}
    \label{fig:assemblies_for_nearly_0_tilts}
\end{figure}

\rev{Presented results demonstrate that even such an elementary modular system based on a single unit provides a substantial flexibility in controlling overall mechanical properties by local rotations of mesostructural units by $90^{\circ}$. The outcomes of experimental tests performed on the additively manufactured specimens confirm that the assembly properties can be accurately designed using a numerical analysis and optimization.}
%Based on the presented results we conjecture that the proposed ``infant'' modular system provides a substantial flexibility in controlling overall mechanical properties by local rotations of mesostructural units by $90^{\circ}$. The outcomes of experimental tests performed on the additively manufactured specimens confirm that the assembly properties can be accurately designed using a numerical analysis \rev{but not predicted simply by a heuristic, analytical or semi-analytical tool}.
%
\begin{figure}[htp]
  \centering	
  \begin{tabular}{ccccc}
    \includegraphics[width=0.17\textwidth]{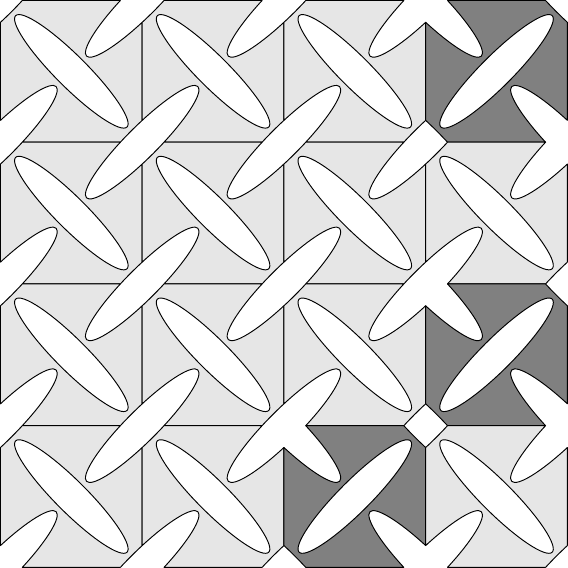} &\includegraphics[width=0.17\textwidth]{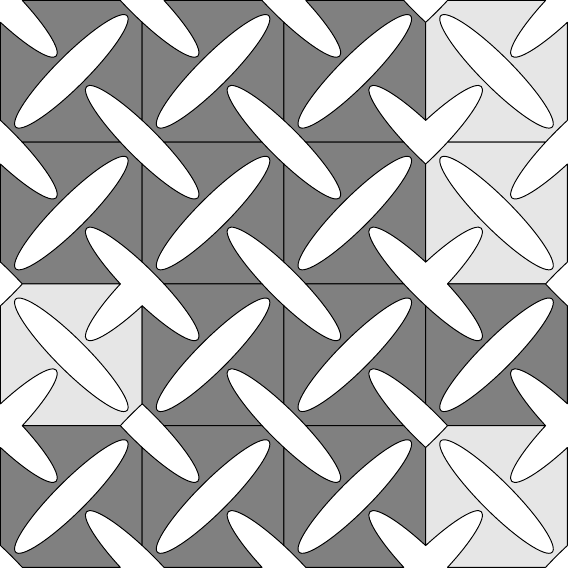} &\includegraphics[width=0.17\textwidth]{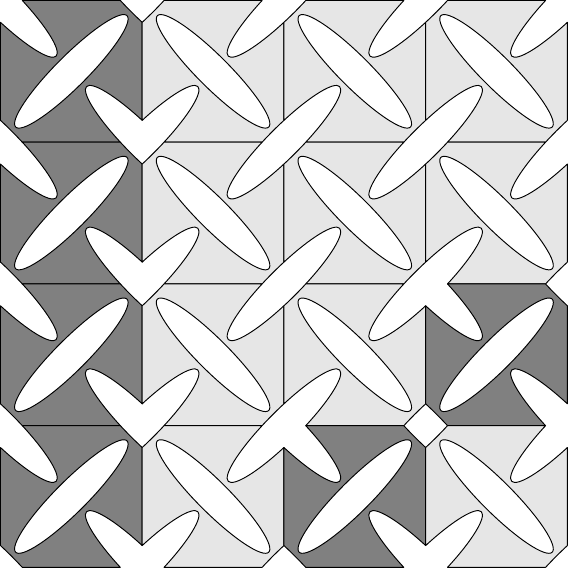} &\includegraphics[width=0.17\textwidth]{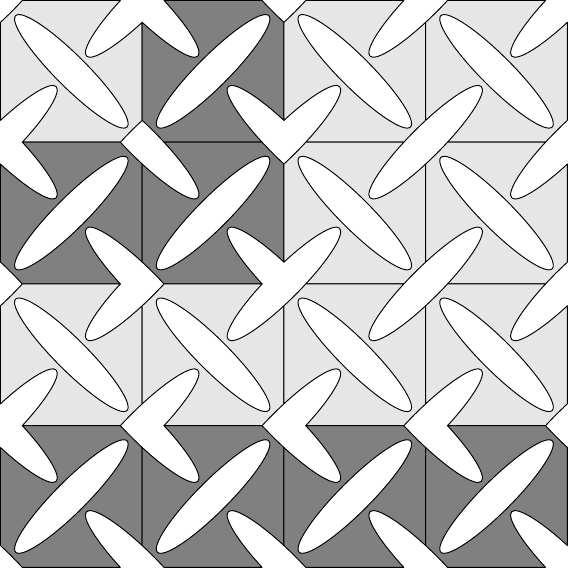} &\includegraphics[width=0.17\textwidth]{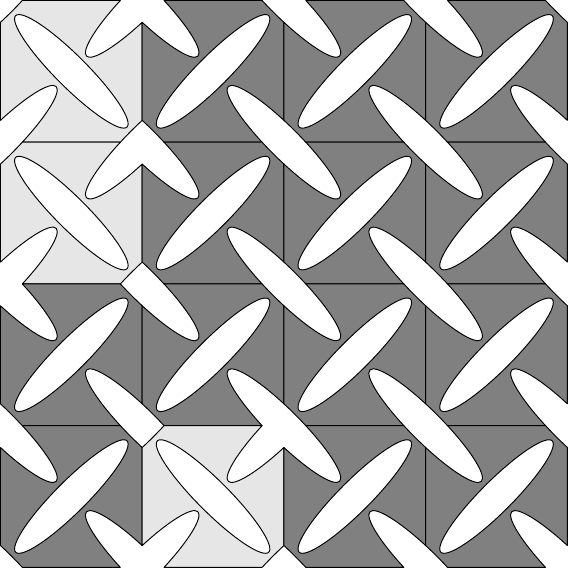}\\
    (a) & (b) &(c) &(d) &(e)
\end{tabular}
  \caption{\rev{Solid assemblies for tilt angles (a) $\phi=-0.100^\circ$, (b) $\phi=0.273^\circ$, (c) $\phi=0.575^\circ$, (d) $\phi=0.912^\circ$, (e) $\phi=1.912^\circ$. Highlighted tiles are rotated by $90^\circ$ with respect to the setting adopted in Figure~\ref{fig:puc_geometry}(a).}}
    \label{fig:assemblies_for_various_tilts}
\end{figure}
\rev{
On the other hand, avoiding numerical analyzes in the design process is difficult, given the emergence of highly-localized instability-like behavior when snapping from one configuration to another (pronounced namely for aggregates with imperfect contacts). Because of this phenomenon, we were unable to devise any simple heuristic, analytical, or semi-analytical tool for predicting the overall assembly response \emph{a priori}. To further highlight this difficulty, Figure~\ref{fig:assemblies_for_various_tilts} shows the intricate, often counter-intuitive dependence of the tilt angle on the tile configurations, whereas Figures~\ref{fig:assemblies_for_nearly_0_tilts} depicts two distinct assemblies delivering nearly the same structural response.
}
%\rev{The latter conjecture is supported by observing the evolution of the assembly plans with respect to the magnitude of the tilt angle $\phi$ as shown in Figure~\ref{fig:assemblies_for_various_tilts}. Such a generalization seem to also difficult to conceive after a careful revision of a series of consecutive assembly plans distinct in a very local detail.}
%
%
%\rev{On the other hand, approaching the problem from the opposite direction, i.e. taking a pair of tilt angles as close as possible, the two assembly plans exhibit some similarity, Figure~\ref{fig:assemblies_for_nearly_0_tilts}. Clearly, the situation is further pronounced when dealing with assemblies with imperfect contacts, as these act as further sub-scale phenomenon that contributes to the buckling-like snaping of the conglomerate behaviour due to local changes in geometry.}

\rev{Nonetheless, to the merit of the proposed modular concept}, it is scalable up to the limits of the manufacturing hardware and extensible to three dimensions. New possibilities in terms of the design flexibility are foreseen by incorporating the concept of Wang tiles~\cite{Wang_1961, Novak_2013, Doskar_2016, Doskar_2018}. As opposed to commonly produced disordered structures, e.g., metal foams, the behavior of the jigsaw puzzle system augmented with Wang tiling would be fully predictable and customizable with respect to specific needs. After translating the concept to practical applications, automated assembly of tiles/modules based on optimized plans is envisaged.

\section*{Acknowledgment}
The support by the Technology Agency of the Czech Republic and Ministry of Industry and Trade under research projects No.~TACR~TH02020420 and MPO~FV10202, respectively, is gratefully acknowledged. \rev{In addition, this work was partly sponsored by CTU grant \\No. SGS18/037/OHK1/1T/11.}

\end{document}